# SmashEx: Smashing SGX Enclaves Using Exceptions

Jinhua Cui*[†]
National University of Defense Technology
Changsha, China
jhcui.gid@gmail.com

Jason Zhijingcheng Yu*
National University of Singapore
Singapore
yu.zhi@comp.nus.edu.sg

Shweta Shinde
ETH Zürich
Zürich, Switzerland
shweta.shivajishinde@inf.ethz.ch

Prateek Saxena
National University of Singapore
Singapore
prateeks@comp.nus.edu.sg

Zhiping Cai
National University of Defense Technology
Changsha, China
zpcai@nudt.edu.cn

## ABSTRACT

Exceptions are a commodity hardware functionality which is central to multi-tasking OSes as well as event-driven user applications. Normally, the OS assists the user application by lifting the semantics of exceptions received from hardware to program-friendly user signals and exception handling interfaces. However, can exception handlers work securely in user enclaves, such as those enabled by Intel SGX, where the OS is *not* trusted by the enclave code?

In this paper, we introduce a new attack called SmashEx which exploits the OS-enclave interface for asynchronous exceptions in SGX. It demonstrates the importance of a fundamental property of safe atomic execution that is required on this interface. In the absence of atomicity, we show that asynchronous exception handling in SGX enclaves is complicated and prone to re-entrancy vulnerabilities. Our attacks do *not* assume any memory errors in the enclave code, side channels, or application-specific logic flaws. We concretely demonstrate exploits that cause arbitrary disclosure of enclave private memory and code-reuse (ROP) attacks in the enclave. We show reliable exploits on two widely-used SGX runtimes, Intel SGX SDK and Microsoft Open Enclave, running OpenSSL and cURL libraries respectively. We tested a total of 14 frameworks, including Intel SGX SDK and Microsoft Open Enclave, 10 of which are vulnerable. We discuss how the vulnerability manifests on both SGX1-based and SGX2-based platforms. We present potential mitigation and long-term defenses for SmashEx.

## 1 INTRODUCTION

Exceptions are a basic functionality available on modern processors and are ubiquitously used by the OS and real-world applications. The OS makes use of exceptions for multiplexing processes and resources, e.g., via timer interrupts and page faults. Applications use programmatic constructs, such as exception and signal handling, to deal with dynamic events or runtime errors. The underlying OS is in charge of monitoring and delivering hardware generated-exceptions to a user process. This design allows application developers to focus on what to do when an event occurs.

Recently, a new form of hardware isolation has been enabled by enclaves such as those provided by Intel SGX. SGX allows user applications to be partitioned into hardware-isolated compartments called enclaves, which are protected from privileged system software (e.g., the hypervisor and the OS). The main guarantee provided by enclaves is protecting the confidentiality and integrity of code running in them. Enclaves are an important step, for example, towards reducing the dependence on privileged OSes and towards confidential computation [3, 4]. This presents a unique security model—a trusted enclave running alongside an untrusted OS. This paper studies how exceptions are handled on SGX, a platform where the OS and user enclave do *not* trust each other.

Exceptions are events that hardware generates and software handles. There are two design choices for enabling exceptions for enclaves. The trusted hardware can directly deliver the exceptions to the enclave code. Alternatively, the hardware can deliver it to the OS, as in non-SGX systems. The current SGX implementation takes the second approach. In such a design, the OS can route an exception to the enclave along with the description of the exception event. Once the exception is delivered to the enclave by either mechanism, the enclave can execute the exception handler. Since the enclave does not trust the OS, this interface requires careful design to ensure security. There are three entities interacting: the user enclave, the trusted SGX hardware, and the untrusted OS.

Many exceptions are synchronous in the sense that the enclave code can control when these exceptions are raised. But, certain exceptions are asynchronous, i.e., they can be triggered outside the control of the enclave. The OS has the power to trigger such exceptions at any time. The design of any hardware enclave abstraction that supports asynchronous exceptions needs to provide at least three security properties on enclave-OS context switches:

- *Register state save/restore.* When an exception interrupts an enclave, the integrity and confidentiality of the requisite enclave register state should be preserved in the presence of a malicious OS.
- *Safe control resumption.* After an exception, the execution resumes either at the point of interruption or at the start of an enclave-defined handler.
- *Atomicity.* The hardware must support sufficient mechanisms for the enclave to prevent exception handling when it is executing inside certain critical sections.

The SGX hardware provides the first two properties, but not the third. An enclave can turn off delivery of certain programmer-defined exception throughout its execution by statically setting its hardware configuration. However, if the enclave does not statically disable exceptions—which is useful for signal handling—SGX does not allow the enclave to selectively mask exceptions at certain times during execution. This effectively means that the SGX hardware does not provide any explicit runtime primitives for ensuring atomicity of critical sections in the enclave, when exceptions are statically enabled. The lack of such a primitive opens up enclaves to *re-entrancy vulnerabilities* which can in turn lead to serious exploits.

To demonstrate this clearly, we introduce a powerful attack called SmashEx, which does not assume any side channels or pre-existing memory safety bugs in the enclave application code. We successfully execute the attack to compromise both confidentiality and integrity guarantees for enclave applications on SGX. Our attack on SSL implementations for instance can cause a malicious OS to spill out secret keys residing in private memory. To demonstrate the full power of SmashEx, we leverage the re-entrancy vulnerability to effect code-reuse (e.g., ROP [52]) and arbitrary memory disclosure attacks on enclaves. We construct end-to-end PoC exploits for two widely-used SGX runtimes: Intel SGX SDK [17] and Microsoft Open Enclave [47]. We target an OpenSSL implementation based on Intel SGX SDK and the cURL application based on Open Enclave respectively. The attacks are demonstrated on the latest SGX2 hardware, but also extends to SGX1 runtimes that have asynchronous exception handling enabled.

In this paper, we explain why the root re-entrancy vulnerability exploited by SmashEx is fundamental—if we want to support asynchronous exception handling on SGX, careful re-entrant design in the enclave is critical. In total, we survey 14 SGX runtime frameworks and deem that the vulnerability affects 10 of them on SGX2. While the exploits do not immediately carry over to 4 of the runtimes, we point out that this comes at the cost of a limit to their exception handling functionality or extra complexity in their design and implementation. We discuss the effectiveness of various software mitigations for SmashEx. We recommend potential hardware abstractions for exposing atomic execution primitives to enclaves to simplify defenses. These may be of independent interest to future enclave designs.

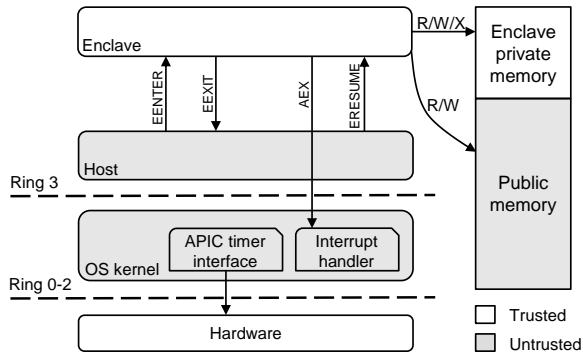

Figure 1: SGX enclave interfaces and memory protection.

**Contribution.** Much prior attention has been devoted to safe data and control exchange at the enclave-OS interface (e.g., for Iago attacks [36]). Our main contribution is to highlight a third missing defense primitive at the enclave-OS interface: ensuring *atomicity* in re-entrant enclave code. When enclaves support the standard programming model of asynchronous exceptions, re-entrancy concerns arise. Our SmashEx attack makes this issue concrete for study.

## 2 BACKGROUND

Intel SGX introduces the notion of *enclaves*—hardware isolated memory regions for sensitive execution. We refer to the code that executes inside an enclave as *enclave software*. The code and the data of enclave software, including its stack (*enclave private stack*), are stored inside enclave memory and protected by the SGX hardware. In the SGX trust model, only the hardware and the enclave software are trusted. All the other software on the system, including privileged software such as the OS, is considered *untrusted*. This includes the user process in charge of creating and interacting with the enclave, which we refer to as its *host process*. The SGX hardware does not allow the untrusted software to access enclave memory. However, enclave software can read or write memory regions outside the enclave boundary, which are also accessible to the host process. We refer to such a shared virtual address space accessible to both an enclave and its host process as the *public memory*.

Enclave software requires mechanisms to request services from the non-enclave/OS code as well as to receive notifications (e.g., signals) from it. The SGX hardware has two kinds of interfaces, synchronous and asynchronous, for switching between the OS and an enclave. Figure 1 depicts such interfaces alongside the protected memory region for an SGX enclave.

**Synchronous Entry/Exits.** Synchronous entry/exits are needed in enclaves to interface with the host application and the OS for synchronous or blocking communication. To help safeguard the interface, the SGX hardware strictly restricts the transfer of control between enclave and non-enclave code. Two specific instructions, EENTER and EEXIT, are used to synchronously enter and exit an enclave respectively. The EENTER instruction jumps to a fixed enclave entry point that is pre-configured during enclave creation.

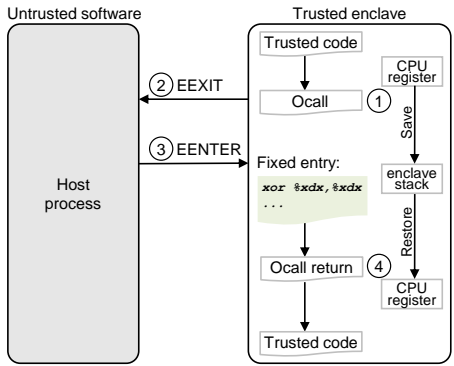

Figure 2: Synchronous entry/exit in an `ocall` on SGX.

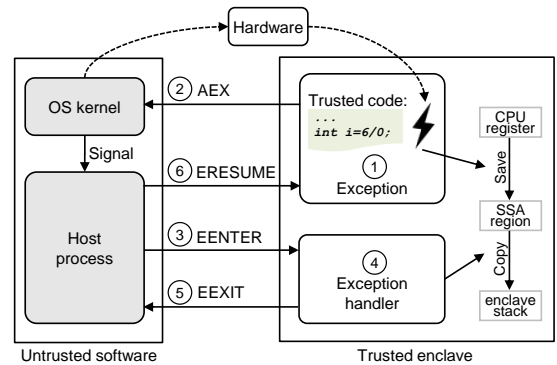

Figure 3: Exception handling mechanism in Intel SGX SDK. The SGX hardware performs an AEX and transfers the control to the OS when an exception occurs (①②). The OS delivers a corresponding signal to the host process, which then re-enters the enclave via EENTER (③). The enclave performs in-enclave exception handling (④) and exits to the host process via EEXIT (⑤), which then resumes the enclave execution via ERESUME (⑥). During this process, the CPU state of the interrupted enclave is first saved into the SSA upon the AEX, from which it is then copied to the enclave private stack during in-enclave exception handling.

An enclave can specify a public memory location and exit to it via the EEXIT instruction. EENTER and EEXIT do *not* scrub or replace the register state during context switches. Instead, the hardware keeps most of the registers unchanged. It is the responsibility of the enclave software to prepare the register state for enclave execution after EENTER and to prevent leaking secrets through the register state after EEXIT. This is necessary for normal functionality, for example, to propagate any data arguments between the enclave and the OS. An enclave can provide functions for untrusted software to invoke in a so-called ecall, which consists of a paired EENTER and a subsequent EEXIT. In addition, synchronous entry/exits can also be used to support ocalls, where enclaves request to invoke functions provided by untrusted software (shown in Figure 2).

Since EEXIT and EENTER do not take care of the register state, the enclave code has to save the enclave CPU context on its private stack and restore it when returning from the ocall later. The ocall interface has been the subject of much scrutiny in prior work, largely due to the risk of Iago attacks [36, 44, 58, 61].

**Asynchronous Entry/Exits.** In addition to synchronous exits, an enclave can exit asynchronously as a result of exceptions (e.g., timer interrupts, page faults, division-by-zero). During such an AEX (Asynchronous Enclave eXit), the enclave stores the current enclave execution context in a special data structure called the *State Save Area (SSA)* located inside the enclave private memory. Asynchronous entry/exits are different from synchronous events because they can arise at any time during the enclave execution, interrupting it involuntarily. To ensure safe enclave-OS transitions, the SGX hardware implements the following mechanisms:

- *Safe control resumption.* At an AEX, the hardware automatically stores the current instruction pointer (rip) in the SSA. The untrusted host process may execute an ERESUME instruction later to transfer control back to the enclave. At this point, SGX hardware enforces that the enclave resumes execution from the rip value stored inside the SSA.
- *Register save/restore.* In addition to rip, the hardware saves the remaining enclave execution context (e.g., general-purpose registers) in the SSA. Before exit, the hardware scrubs the register values to prevent data leakage through them. On ERESUME, the hardware restores the register values from SSA.

**Asynchronous Exception Handling in SGX Enclaves.** The simple mechanisms above are sufficient to protect an enclave while allowing exceptions to interrupt its execution. However, in order to also allow the enclave to handle exceptions (including deciding the resumption point by modifying the SSA content), a more complex mechanism is designed (shown in Figure 3) in SGX runtimes. Instead of resuming the enclave immediately via ERESUME, the untrusted host process re-enters the enclave using EENTER and passes relevant information about the exception. Note that this new flow starts with a normal EENTER which leaves the rsp value uninitialized, so the enclave has to set up its private stack before executing any real exception handler. In both Intel SGX SDK and Microsoft Open Enclave (among others; see Section 8), the enclave loads the stack pointer from the saved rsp in the SSA, effectively reusing the same stack of the interrupted thread. After the enclave finishes handling the exception in the SDK, it uses EEXIT to return control to the untrusted software, which then resumes the enclave execution via ERESUME.

**Key Observation.** To perform exception handling, the enclave needs to be re-entered and operate on a context that overlaps with that of the interrupted thread. Therefore, the above design for in-enclave handlers requires the enclave to be *re-entrant*.

**SGX Runtimes.** Since the SGX enclave programming model is significantly different from a traditional one, it can be cumbersome for developers to use the low-level SGX interfaces. Therefore, enclave developers usually use frameworks that provide high-level abstractions to hide away the details of exception handling, ocall, ecall, and so on. Such software frameworks, referred to as SGX enclave runtimes in this paper, execute inside enclaves. Since runtimes are a part of the enclave trusted computing base, their design and implementation are crucial to the security of enclave applications. We survey 14 runtimes (Table 1) and find that they have varying degrees of exception handling support.

| Software | Version | Vulnerable to SmashEx? | Exception handling |
|---|---|---|---|
| **Intel SGX SDK** [17] | 2.13 | ● | ● |
| **Microsoft Open Enclave** [48] | 0.15.0 | ● | ● |
| **RedHat Enarx** [13] | 02dab73 | ● | ● |
| **Graphene-SGX** [35] | 1.1 | ○ | ● |
| **Apache Teaclave** [1] | 0.2.0 | ● | ● |
| **Google Asylo** [31] | 0.6.2 | ● | ● |
| **Fortanix Rust EDP** [41] | 3341ce1 | ○ | ○ |
| **Alibaba Inclavare** [15] | 0.6.0 | ○ | ○ |
| **Ratel** [26] | 1.1 | ○ | ● |
| **SGX-LKL** [51] | b6e838e | ● | ● |
| **EdgelessRT** [12] | 8a6f11f | ● | ● |
| **Rust SGX SDK** [64] | 1.1.3 | ● | ● |
| **CoSMIX** [50] | 4e67f55 | ● | ● |
| **Veracruz** [28] | cbf01a9 | ● | ● |

Table 1: Summary of different enclave runtime designs. ● denotes that the enclave runtime is deemed exploitable by SmashEx and ○ denotes that the enclave runtime is deemed unexploitable by SmashEx under any enclave settings. For the exception handling, ● denotes handling asynchronous exceptions is supported in enclaves and ○ denotes no support for handling exceptions in enclaves.

## 3 ATTACK OVERVIEW

An enclave must be re-entrant to safely handle exceptions by itself. However, an important primitive, namely *atomicity*, is missing. We illustrate the need for atomicity by outlining SmashEx, a novel attack that exploits a re-entrancy vulnerability present in many SGX frameworks.

### 3.1 Threat Model

**Assumptions.** The security-sensitive part of the application executes, together with any runtime libraries, in the victim enclave. The enclave interacts with the external environment, including the OS and the host process, which are assumed to be arbitrarily malicious. The SGX hardware is trusted. The enclave code is assumed to be benign and we do *not* assume that it has any auxiliary bugs or side channels that aid the attacker. We assume that the enclave code, however, is running without ASLR (Address Space Layout Randomization), i.e., certain critical addresses are deterministic and known to the attacker. This is the default setup on nearly all SGX platform runtimes we have surveyed. We discuss how to extend our attacks when ASLR or other auxiliary defenses are enabled in Section 9. The enclave is assumed to have enabled the asynchronous exception handling interface.

**Attack Goal & Scope.** Our attack goal is to break the basic memory protection guarantees offered by SGX enclaves. Specifically, we aim to break confidentiality by enabling an arbitrary memory disclosure attack by which the attacker can reveal the victim enclave memory contents in full. To break integrity, our attack aims to enable ROP attacks in the enclave. Most runtimes that support in-enclave exception handling are susceptible to our attack. In our survey summarized in Table 1, 12/14 runtimes support this functionality. Among the 12 runtimes, we find that 10 are susceptible to the SmashEx attack. We present proof-of-concept (PoC) exploits for two of the most popular runtimes, Intel SGX SDK and Open Enclave, as case studies in Section 7. Intel SGX SDK is widely used in multiple other runtimes and Open Enclave is part of the Microsoft Confidential Computing Framework (CCF). Since SmashEx arises directly on the enclave-OS interface, any application that uses a vulnerable runtime is exploitable. We have also constructed PoC exploits for all but RedHat Enarx, for which we have confirmed the exploitability through code inspection (see Section 8).

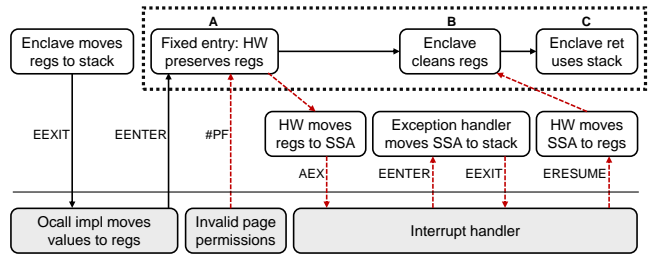

Figure 4: State diagram depicting the re-entrancy vulnerability. The clear boxes denote enclave states, and the gray boxes denote OS states. The solid black arrows show the `ocall` execution flow where the dotted black box denotes the critical section. The dashed red arrows show the in-enclave exception handling flow, injected by the OS when the enclave is in critical section, thus corrupting the enclave state (i.e., stack).

### 3.2 The Re-entrancy Vulnerability

To demonstrate the re-entrancy issue clearly, we outline the flow of exception handling on Intel SGX SDK for SGX2, both under normal execution and under a SmashEx attack. It executes similarly on SGX1 and extends to other runtimes (see Section 8).

Consider the flow that handles returning from an `ocall`. Figure 4 shows this flow with black solid arrows inside the dotted box, which executes logic labeled $A \rightarrow B \rightarrow C$ in that sequence. This flow, however, can be interrupted when asynchronous exceptions arrive. The dashed red arrows in Figure 4 show the execution flow when handling in-enclave exceptions corresponding to the EENTER → EEXIT → ERESUME previously highlighted in Section 2. Both flows are benign, but operate on overlapping enclave contexts. This clearly highlights that such `ocall` return flow and the exception handling flow should be written with care to ensure that they interleave safely. Specifically, when the `ocall` return flow is interrupted, the enclave should be in a consistent state for the exception handling flow to progress correctly, and when the exception handling flow completes, the enclave state should also be ready for the enclave to resume. This adds considerable complexity when in-enclave exceptions are to be supported, regardless of whether the OS is acting maliciously.

In this example, the main vulnerability point that enables SmashEx is in the `ocall` return flow, which requires *atomicity* for executing certain critical sections that update the enclave context (shown in the dotted black box in Figure 4). The state transitions for the `ocall` return flow must check and clean up the register values returned by the OS and then use them to set up the enclave private stack. Before this is finished, the enclave is in an inconsistent state, with register values (in particular, the stack pointer rsp) provided by the untrusted OS.

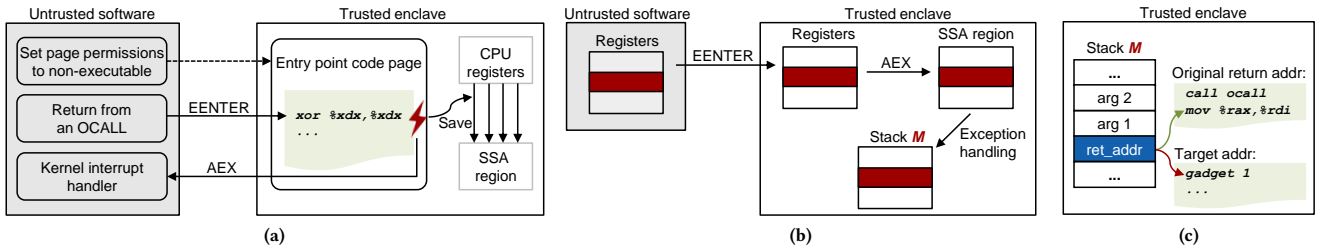

Figure 5: Sub-components of SmashEx. (a) Injecting an AEX at the fixed entry point; (b) Corrupting the in-enclave memory (corrupted data is depicted in red); (c) Returning to the target that the attacker has specified in the controlled anchor (in blue).

It is important, therefore, that the enclave should not be interrupted to perform exception handling when it is still executing in such an inconsistent state. However, the SGX enclave abstraction does not provide primitives for ensuring atomic execution of critical sections in the enclave. For example, it is not possible to selectively mask interrupts for certain critical sections. The enclave must either statically disable all user-defined asynchronous exception handling[1] or risk being interrupted arbitrarily if exception handling is statically allowed. The lack of atomicity results in a powerful attack vector which SmashEx leverages.

## 3.3 High-level Attack Steps

As illustrated in Figure 5, SmashEx starts with the attacker triggering an exception immediately after the enclave is entered (via EENTER) to return from an ocall (Figure 5a). The hardware copies the attacker-controlled registers into the SSA region, which the enclave exception handler in turn uses to determine the stack address and the data to later use. This gives the attacker the capability of corrupting the stack content of the enclave (Figure 5b). By carefully crafting the register values, the attacker can exploit this capability to corrupt an enclave stack location that the enclave will later use to load a return address (we call this location an *anchor*), thereby hijacking the control flow of the enclave (Figure 5c). Figure 6 describes the detailed steps SmashEx follows, which we will discuss in Sections 4, 5, and 6.

## 3.4 Difference to Prior Attacks

SmashEx is the first attack that demonstrates the exception handling attack vector in Intel SGX. that SmashEx is conceptually close to known prior attacks on enclave memory safety, synchronization, and scheduling. However, SmashEx is significantly different. Briefly, previous attacks assume much more than SmashEx:

- AsyncShock [2] assumes that the synchronization logic (e.g., using mutexes) between two or more enclave threads is buggy. In contrast, the vulnerability SmashEx exploits arises due to atomicity violation in the enclave-OS interaction, to which thread synchronization is irrelevant.

---
[1]In SGX, the runtime can disable all in-enclave exception handling by setting the TCS.NSSA enclave configuration to one.

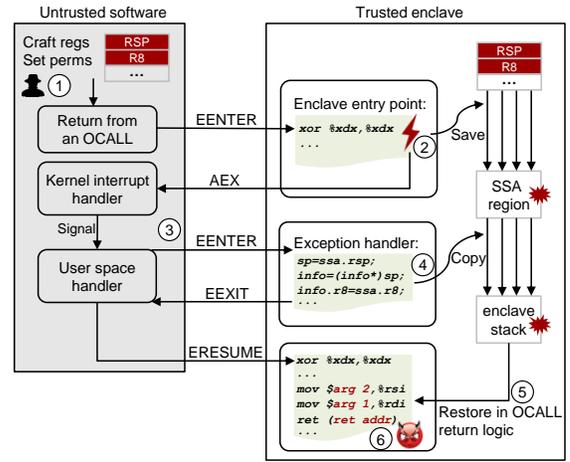

Figure 6: SmashEx Overview. The attacker directly controls the untrusted software, including the OS and the host process. Data that the attacker can control is in red.

- Game of Threads [63] shows that by manipulating thread scheduling, a malicious OS is able to stably exploit faulty thread synchronization logic in enclave applications. For example, for machine learning training workloads without frequent thread synchronization, malicious scheduling can degrade accuracy and bias the model.
- The Guard's Dilemma [33] assumes the existence of memory vulnerabilities in the enclave application and uses them to demonstrate code-reuse attacks such as ROP. In SmashEx, we do not assume any such pre-existing memory errors.

## 4 ARBITRARY WRITE CAPABILITY

The SmashEx attack starts with enabling the attacker to perform an arbitrary write to an attacker-specified anchor location (Figure 5a).

**Step 1: Preparing Malicious Register Values.** The attacker loads malicious values into registers right before EENTER to ensure that the attacker-specified register state is preserved.

**Step 2: Injecting an Exception at the Precise Time/Location.** SmashEx requires that the AEX event occurs shortly after enclave entry, before the enclave cleans up the register state. There are at least two ways to achieve this:

*(a) Page faults.* We identify the page that contains the first enclave instruction and pre-set its access permissions to be non-executable. The malicious OS is still in charge of the enclave page permissions and can trivially do this. In particular, the OS knows the exact page address because it has to set up the enclave memory layout before launching an enclave. Note that this address cannot be hidden from the attacker (e.g., via randomization [33, 54]) because the hardware has to know the exact entry point for the enclave. As described in Section 3, we use this mechanism to trigger a page fault when the enclave attempts to execute the first enclave instruction. When the page fault is triggered, the hardware delivers it to the attacker-controlled OS, which then forwards it to the enclave for handling.

*(b) Timer interrupts.* We can also interrupt the enclave execution via the APIC timer. Prior work [62] has shown that the OS can invoke the APIC timer interface to precisely interrupt the enclave execution at any desired point. The remaining challenge is to inject the timer interrupt at the exact moment. Before returning from an ocall, we set the APIC timer to the one-shot mode. We set the timer count such that the interrupt will occur immediately after EENTER. In order to stably achieve this, we execute the enclave in debug mode and tune the timer count to interrupt at the right time. In our experiments, when we reuse this same timer count in production mode, we can reliably inject the interrupt.

**Step 3: Re-entering the Enclave for Exception Handling.** After the untrusted OS gains control because of the AEX, it re-enters the enclave (via EENTER) for handling the exception. This time, the attacker allows the enclave to progress after the enclave entry by reverting the access protection to the original permissions, if page faults are used in Step 2.

**Step 4: Tricking the Enclave into Using Malicious Values.** To handle the AEX, the enclave first needs to prepare the enclave stack for the in-enclave handler by loading the rsp register from the SSA (see Section 2). It then copies the SSA content onto this stack for the handler to use as function arguments. Since the attacker can control the values of rsp and the other saved registers in the SSA, it has gained the capability of tricking the enclave into writing an attacker-specified value to an attacker-specified stack location. To hijack the enclave control flow, the attacker uses this capability to control an anchor, i.e., a location later used by the enclave to retrieve a return address. More specifically, the attacker can choose the stack location that stores the return address for the current ocall as the anchor.

**Achieved Capability.** Using the above steps, the attacker has the capability to write an arbitrary value to a particular location. This write capability is the first part for scaffolding a control-flow hijacking attack. In Sections 5 and 6, we explain how to leverage this capability to effect powerful end-to-end attacks.

## 5 SETTING UP THE STACK

So far, the attack has only corrupted one anchor location on the stack. Our final goal is to demonstrate a powerful ROP attack [52]. To this end, our next step is to escalate the attacker's capability to:

- Point the stack pointer to an attacker-controlled region;
- Trick the enclave into using the value stored at the anchor for a control-flow transfer.

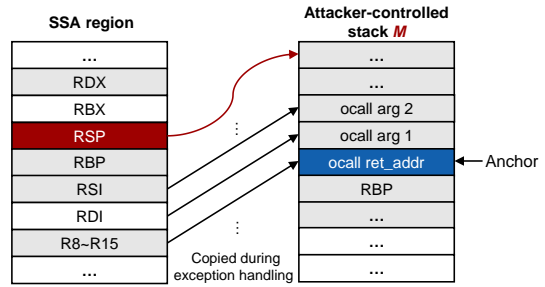

**Figure 7: Enclave memory corruption by SMASHEX during exception handling in Intel SGX SDK. The attacker is able to control the stack pointer in the SSA (in red) and the anchor on the enclave private stack (in blue). The other data that the attacker can control in SMASHEX is shown in gray.**

**Step 5: Pointing the Stack Pointer to Attacker Memory.** A ROP attack requires the attacker to control data on the victim's stack so gadgets can be strung together through a series of return addresses. As illustrated in Figure 7, in the exception handling process (Section 2), the in-enclave exception handler copies the register state stored in the SSA region into a region of the enclave private stack (say $M$). This SSA state consists of a group of registers which the SGX hardware has saved during the AEX event, and which, as explained earlier, are attacker-controlled. The attacker may therefore use $M$ to store the gadget addresses. To make the enclave use this region as its stack, the attacker needs to point the enclave rsp value to it *after* the ocall return is completed. This can be easily done by setting the anchor value to the address of a gadget that moves the value of a register (which the attacker already controls through returning from ocall) into rsp.

SMASHEX does not require the memory region for preparing the gadget addresses to be the same as the region $M$. Since the SGX hardware allows an enclave to use a stack inside the public memory, the attacker can simply set up the gadget addresses in a buffer in the host process (located in the public memory), and use the same method to point rsp to it. We use this strategy for our exploit on Open Enclave and the earlier one for Intel SGX SDK.

**Step 6: Effecting a Control Transfer Using the Anchor.** After Step 5, the region $M$ itself is the stack with attacker-controlled values and the stack pointer (rsp) points to it. This is part of what we need to start a ROP attack. It remains to cause a control-flow transfer with the corrupted anchor value (Figure 5c). The exception handler does *not* immediately use the anchor value after the copy logic. Several control transfers and context switches[2] happen before the anchor is used, but the content of $M$ remains unchanged.

Note that although it is possible to set the anchor to any value, pointing it to the public memory will merely crash the enclave (hence falling short of a code-reuse attack) since SGX enclaves cannot execute code from public memory. For this reason, we must confine the anchor value to private memory addresses.

---

[2]The exception handler performs several other operations and exits the enclave using an EEXIT instruction to the untrusted code, which then performs an ERESUME instruction to transfer control back to the enclave to complete the exception handling and the ocall return flow interrupted. This part of the logic uses the anchor in a control-transfer instruction—this is why we chose the anchor to be the return address used in resuming after the last ocall.

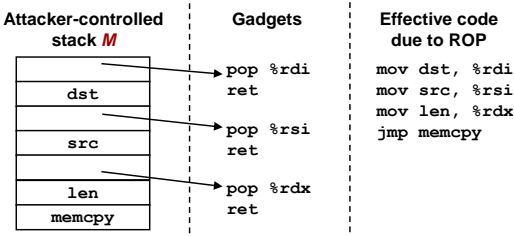

**Figure 8: A chain of ROP gadgets in the malicious stack to invoke `memcpy` with attacker-controlled arguments.**

**Achieved Capability.** At the end of Steps 5 and 6, the enclave starts executing with the stack content controlled by the attacker. By carefully crafting the stack content, the attacker is able to convert this capability to a full-blown ROP exploit.

## 6 ROP EXPLOITS

At this point, the attacker can already control both the enclave instruction pointer (`rip`) and the enclave stack content. Next, we escalate the attacker's capability to being able to execute a sequence of ROP gadgets that exist in the enclave code [52]. We discuss the ways to achieve this for two different goals: to steal enclave secrets and to execute desirable ROP gadgets.

**Goal 1. Compromising Enclave Confidentiality.** Enclave runtimes (e.g., Intel SGX SDK, Open Enclave SDK) usually implement their own `memcpy` function for in-enclave operations. Such a function performs memory copy on any accessible memory location regardless of the enclave boundary, and accepts three arguments that specify the source and the destination as well as the size of the data to copy. The three arguments are passed in registers `rdi`, `rsi`, and `rdx`. We can use this function in our chain of gadgets. First, we set up arbitrary values into registers using memory-to-register move gadgets. Then we chain a gadget to invoke the `memcpy` function. This allows us to move arbitrary regions of memory to arbitrary locations. For example, we can point the source address argument to the start of the enclave and destination address to a public memory region. Such a gadget will dump the entire enclave memory. Alternatively, we can point the source to sensitive data (e.g., SSL keys, enclave ephemeral keys) to selectively leak secrets. To compromise enclave confidentiality, we use ROP gadgets to manipulate the enclave to execute a `memcpy` function. Through manual inspection, we find and locate `memcpy` implementations in the trusted runtime code of both Intel SGX SDK and Microsoft Open Enclave. We also find three `pop reg; ret` gadgets in the runtimes that allow the attacker to populate the three registers with values from the attacker-controlled external stack. As illustrated in Figure 8, the attacker can perform the desired `memcpy` and steal enclave secrets by chaining the ROP gadgets on the external stack.

**Goal 2. Arbitrary Code Injection in Enclave.** Similarly, the attacker can copy arbitrary code to the enclave memory and execute it. It points the source to a malicious code payload outside the enclave and the destination to an enclave page. Specifically, if the attacker sets the source to a prepared shellcode in the public memory and the destination to an executable and writable region in the enclave private memory, it will be able to inject the shellcode to the enclave. The attacker can then point the subsequent return address on the external stack to the injected shellcode in the enclave to execute it. Since certain applications such as JIT compilers need executable and writable enclave memory regions, SMASHEx can inject and execute arbitrary code. On an SGX2 platform where dynamic adjustment of enclave page permissions is supported, SMASHEx can use ROP gadgets to make certain pages executable and writable before injecting and executing the shellcode.

**Other Desirable ROP Gadgets.** In addition to the ROP gadget chains discussed above, the attacker may also use others that serve a wide range of goals. Prior work [33] has concluded that a ROP attack in an SGX enclave can be very expressive. More specifically, special gadgets available in an SGX runtime (e.g., Intel SGX SDK) enable the attacker to control the entire register file if it already controls `rsp`, `rdi` and `rsi`. SMASHEx meets this criterion required by previous attacks. We can therefore reproduce any attacks shown by this prior work[3] but without their assumptions of common memory vulnerabilities such as buffer overflow in the enclave code.

## 7 ATTACKING REAL SYSTEMS

We present the low-level implementation challenges in executing end-to-end attacks on two applications as case studies: an OpenSSL port with Intel SGX SDK and a cURL port with Open Enclave.

We run all the victim enclaves and the SMASHEx exploits on an Intel NUC Kit NUC7PJYH with SGX2 support, 8 GB DRAM, 128 MB EPC, and a Ubuntu 18.04 installation (Linux kernel 5.4.0-72). For SGX enclave runtimes, we use Intel SGX SDK 2.13, SGX driver 2.11, and Open Enclave 0.15.0 [48], since these are the latest versions available at the time of our experiments.

### 7.1 Intel SGX SDK

**Case Study: OpenSSL v1.1.1i.** Intel SGX SSL [18] is a cryptographic library that uses OpenSSL [23] to provide general-purpose cryptographic services (e.g., key generation, encryption/decryption operations, decision-making statements) for SGX enclave applications. For our end-to-end attack, we target a test program bundled with Intel SGX SSL, where the enclave generates a public/private RSA key pair. By leaking this private key, we show that SMASHEx can breach the Intel SGX protections.

We have to locate the target in-enclave secret before we can launch SMASHEx. For this purpose, we disable ASLR system-wide and pre-run the enclave once to record the addresses. In our attack run, we wait for the enclave to invoke a specific `ocall` that reports the result to the user after finishing the computation. We choose to start our attack after this `ocall`, because by this point, the enclave has created the private key in its private memory. To copy the secret key to the public memory, we use the `memcpy` gadget chain described in Section 6. The attack causes the enclave to copy the 1024-bit key to the public memory.

---

[3]For example, the attacker can chain the `asm_oret` and `continue_execution` gadgets in Intel SGX SDK or `oe_longjmp` and `oe_continue_execution` in Open Enclave to control a wider range of registers [33].

**Implementation Challenges.** We encountered the following two main challenges when launching SmashEx against Intel SGX SDK.

*Bypassing overrun and alignment checks.* During exception handling, the SGX runtime performs security checks to sanitize and ensure consistency of certain enclave states. For instance, the in-enclave exception handler (see Listing 1) derives the stack pointer from the SSA. Then it checks that the stack pointer is a valid enclave stack address and satisfies a pre-defined alignment requirement.[4] We set the malicious stack pointer to a legitimate enclave stack address that obeys the required alignment to pass those checks.

```
1 ... // check validity of thread_data, tcs, stack canary, enclave state,
       exception flag, ssa region
2 ssa_gpr = reinterpret_cast<ssa_gpr_t *>(thread_data->first_ssa_gpr);
3 sp = ssa_gpr->REG(sp);
4 ... // check stack overrun
5 info = (sgx_exception_info_t *)sp;
6 if(ssa_gpr->exit_info.valid != 1) { // exception handlers are not allowed to
       call in a non-exception state
7     goto default_handler;
8 }
9 ...
10 info->cpu_context.r8 = ssa_gpr->r8;
11 ...
12 info->cpu_context.r15 = ssa_gpr->r15;
13 ... // alignment will be checked after exception is handled
```

**Listing 1: Operations and security checks during exception handling in Intel SGX SDK.**

The `ocall` return logic (see Listing 2) also includes important checks that our attack has to circumvent. For instance, before restoring the `ocall` context,[5] the enclave sanitizes the `ocall` context pointer to ensure that it is on the enclave stack (Lines 3 and 5). In addition, the enclave checks the validity of part of the `ocall` context content (Lines 7 and 9). However, those checks do not cover the data that SmashEx needs to overwrite. We craft a legitimate stack pointer value to control the anchor without corrupting the checked memory region. In this way, our attack bypasses the checks.

```
1 uintptr_t last_sp = thread_data->last_sp;
2 ocall_context_t *context = reinterpret_cast<ocall_context_t*>(thread_data->
       last_sp);
3 if(0 == last_sp || last_sp <= (uintptr_t)&context)
4     return SGX_ERROR_UNEXPECTED;
5 if(last_sp > thread_data->stack_base_addr - 30 * sizeof(size_t))
6     return SGX_ERROR_UNEXPECTED;
7 if(context->ocall_flag != ocall_flag)
8     return SGX_ERROR_UNEXPECTED;
9 if(context->pre_last_sp > thread_data->stack_base_addr ||
10     context->pre_last_sp <= (uintptr_t)context)
11     return SGX_ERROR_UNEXPECTED;
12 thread_data->last_sp = context->pre_last_sp;
13 asm_oret(last_sp, ms);
```

**Listing 2: Security checks before restoring the `ocall` context in Intel SGX SDK.**

*Restoring host process stack after AEX.* Recall that in Step 1 of SmashEx, we prepare the `rsp` with a malicious address that points to the enclave private memory. When we trigger an AEX in Step 2, the hardware retains the `rsp` even after exiting the enclave. Additionally, the hardware transfers control to the OS for kernel exception handling. The kernel generates a corresponding signal for the exception and wants to deliver the signal to the host process. The kernel attempts to do this by using the `rsp` to place the signal-related information on the host process stack. Since the `rsp` still points to an in-enclave address, the kernel cannot perform this operation. However, for Step 3 of our SmashEx, it is necessary that the attacker handle this signal in the host process. We ensure that when the OS accesses the `rsp` it is pointing to a host stack location with the `sigaltstack()` system call, which allows a user process to specify a separate signal handling stack. Alternatively, when the attacker moves malicious values to `rsp` in Step 1, we can save the current `rsp` in the host process. After the AEX, when the control comes to the kernel, we restore the saved `rsp` value to the `rsp` register. Note that the malicious `rsp` value has already been stored in the SSA at this point. Therefore, we can safely change the `rsp`.

---

[4]In Intel SGX SDK 2.13, the `is_valid_sp` function performs such checks.
[5]ocall context is a data structure on the enclave stack that stores the context of the enclave before an ocall.

Our PoC integrates those two mechanisms to overcome the implementation quirks of Intel SGX SDK.

## 7.2 Open Enclave SDK

**Case Study: cURL v7.67.1.** The cURL library implements a wide range of application-layer network protocols, including HTTP, HTTPS, SMTP, and so on. Open Enclave provides an official port of cURL [49] to allow applications that require secure network protocols (e.g., HTTPS) to benefit from the protection of SGX. The enclave private memory contains several pieces of sensitive information such as secure channel keys, enclave private keys, and HTTPS plaintext responses.

We run SmashEx on an Open Enclave cURL test program and dump the whole enclave private memory to the public memory. This will allow us to extract all the secrets inside the enclave private memory. We obtain the virtual address ranges of the enclave private memory regions by consulting the untrusted library of Open Enclave. The library is responsible for creating the enclave, and is therefore aware of the enclave address space layout. To ensure that the enclave private memory contains secret data at the time of our attack, we wait for the enclave to finish sufficient `ocalls` before launching the attack. In our experiment, we wait until right after the 150*th* `ocall` to start SmashEx, where we use the gadgets from Section 6 to dump the enclave content. The dumped data in our experiment includes secrets such as plaintext HTTPS responses.

**Configuring the APIC Timer.** We use the APIC timer to trigger the AEX at the precise moment [62]. Typically, only the OS kernel can configure the APIC timer. However, in order to trigger the exception at precisely the first instruction of the enclave, we want to shorten the time gap between configuring the APIC timer and entering the enclave. Therefore, instead of configuring the APIC timer inside the kernel space, we map the interface of the APIC timer (memory-mapped I/O) directly to the address space of the host process, and configure the APIC timer in the user space shortly before entering the enclave via EENTER.

## 8 ATTACK EXTENSIBILITY

The ramifications of unsafe re-entrancy in enclave handlers go beyond the target platform configurations—hardware version and runtime—we used for our end-to-end attacks.

## 8.1 Extensibility to SGX1

SGX1 and SGX2 have the same exception handling mechanism. The main difference is that in SGX2, the enclave can request the hardware to notify the enclave about certain exceptions, such as page faults. This allows enclaves to dynamically manage memory.

The attack steps described so far assume SGX2, but they largely also apply to SGX1. Unlike SGX2, SGX1 does not support reporting page faults to the enclave. When such an event occurs, the SGX1 hardware performs an AEX, but with one difference to SGX2: it does so without setting SSA.EXITINFO.valid, a field in the SSA region, to 1. Both Intel SGX SDK and Open Enclave perform a validity check on this field in their exception handlers and only execute the handler if SSA.EXITINFO.valid is 1. In Open Enclave, by the time this check is done, SmashEx can already corrupt the anchor. Therefore, SmashEx works on Open Enclave with SGX1. In Intel SGX SDK, SmashEx needs to bypass the above check to succeed. We examined the 8 exceptions (unconditionally supported exceptions [16]) supported in SGX1. None can be used to trigger the AEX at the enclave entry. As a result, we are not able to exploit Intel SGX SDK on SGX1 with SmashEx.

However, this safety comes with a trade-off in functionality for Intel SGX SDK: with this check in force, Intel SGX SDK disables asynchronous events on SGX1 and does not support programming primitives for user-defined signal handlers. The root vulnerability (i.e., the lack of atomicity) is fundamental. We hypothesized that it would affect SGX1 if Intel SGX SDK allowed execution of in-enclave exception handlers, and confirmed our hypothesis by removing the validity check in the Intel SGX SDK and repeating our attack. Our PoC works successfully on Intel SDK for SGX1 with the one-line validity check removed.

## 8.2 Extensibility to Other Enclave Runtimes

Apart from Intel SGX SDK and Open Enclave, we survey 12 other enclave runtimes to understand how the vulnerability impacts them. We report that 8 of them are vulnerable to SmashEx, and have verified this by constructing SmashEx PoC exploits against them.

**Derivatives of Intel SGX SDK & Open Enclave.** In our survey, 8 enclave runtimes are based on either Intel SGX SDK or Open Enclave SDK. Among them, 6 runtimes use Intel SGX SDK or Open Enclave as it is, without any modification to the exception handling logic. Those include Apache Teaclave [1], Rust SGX SDK [64], CoSMIX [50], and Veracruz [28] which are based on Intel SGX SDK, and SGX-LKL [51] and EdgelessRT [12] which are based on Open Enclave. Since all the relevant interfaces are still exposed and unchanged, such runtimes inherit the vulnerability from the runtime they are based on. The other 2 runtimes, Google Asylo [31] and Ratel [26], use modified Intel SGX SDK. They have altered the behaviors of exception handling or other enclave interfaces relevant to SmashEx and hence need to be examined individually. Google Asylo [31] keeps the original exception handling interface and as a result is vulnerable to SmashEx. However, it also provides an alternative exception handling interface which uses a dedicated stack and cannot be exploited by SmashEx. Ratel [26] is immune to SmashEx because it uses a separate pre-allocated enclave stack for exception handling. We discuss the dedicated-stack design in details in Section 9.1.

**Independent Runtimes.** RedHat Enarx [13] has its own SGX runtime independent of Intel SGX SDK and Open Enclave. Listing 3 shows how it sets up the exception handler stack shortly after the enclave is re-entered for exception handling. Similarly to its counterpart in Open Enclave, the code loads the saved rsp register from the SSA region, shifts it by a fixed offset, and starts storing untrusted register values at that location. We therefore conclude, through our best-effort code inspection, that SmashEx would work successfully on RedHat Enarx. Though open-source, RedHat Enarx does not have fully functioning code base yet [14]. Thus, we were not able to experimentally demonstrate and confirm that SmashEx works on it.

```
1   shl     $12,        %rax        # %rax = CSSA * 4096
2   mov     %rcx,       %r11        # %r11 = &Layout
3   add     %rax,       %r11        # %r11 = &aex[CSSA - 1]
4
5   mov     RSP(%r11),  %r10        # %r10 = aex[CSSA - 1].gpr.rsp
6   sub     $128,       %r10        # Skip the red zone
7   and     $~0xf,      %r10        # Align
8
9   mov     SRSP(%r11), %rax        # %rax = syscall return stack pointer
10
11  xchg    %r10,       %rsp        # Swap to trusted stack
12  pushq   $0                      # Align stack
13  push    %r10                    # Save untrusted %rsp
14  savep                           # Save untrusted preserved registers
```

**Listing 3: Exception handler stack setup in RedHat Enarx.**

The other three runtimes developed independently of Intel SGX SDK and Open Enclave SDK—Alibaba Inclavare [15], Fortanix Rust EDP [41], and Graphene-SGX [35, 60]—are deemed immune to SmashEx through manual inspection. Alibaba Inclavare [15] and Fortanix Rust EDP [41] both simply disable all in-enclave exception handling, which limits the enclave functionality. Graphene-SGX [35, 60] introduces software-based atomicity to safely handle exceptions, which we elaborate on in Section 9.3.

## 9 DEFENDING AGAINST SMASHEX

The proof-of-concept exploits for SmashEx are viable because the enclave runtimes (a) use the common program stack for exception handling; and (b) lack software- or hardware-enforced atomicity. An ideal solution would be to defeat both (a) and (b). However, we discuss the mitigations for these two issues separately. We summarize how certain design choices render enclave runtimes immune to SmashEx, by disabling either requirement (a) or (b):

- Use a dedicated stack for exception handling (e.g., Ratel [26] and the alternative mechanism in Google Asylo [31]);
- Disable exception handling (e.g., Fortanix Rust EDP [41] and Alibaba Inclavare [15]);
- Program the exception handler in a re-entrant way (e.g., Graphene-SGX [35]).

However, those designs come with significant downsides either by limiting the enclave functionality or by introducing complexity.

## 9.1 Dedicated Exception Handler Stack

Unlike the original exception handler in Intel SGX SDK, the exception handling interfaces in Google Asylo and Ratel use a dedicated stack separate from the one used by the interrupted thread. They therefore avoid relying on the rsp value in the SSA region which SmashEx exploits to control the anchor. However, since both of them reserve only one separate stack, exception handling through such interfaces cannot be nested. In other words, if during the handling of an exception, another exception occurs, this new exception cannot be handled inside the enclave. This limits the compatibility between Google Asylo or Ratel and traditional programming models where signals can be nested. One can adapt these runtimes to support nested exceptions by reserving an individual stack for each level of nested exceptions. However, the fixed memory size required by each reserved stack may limit its scalability.

### 9.2 Disabling Exception Handling

Both Fortanix Rust EDP [41] and Alibaba Inclavare [15] are immune to SmashEx because they do not support any in-enclave exception handling. They configure enclaves so that the SGX hardware forbids the untrusted software from re-entering the enclave via EENTER after an AEX. Specifically, the configuration parameter, TCS.NSSA, when set to 1, implies that the hardware can store at most 1 AEX context inside the SSA at any time. Whenever the untrusted software attempts to re-enter the enclave via EENTER following an AEX, the hardware disallows it because of the insufficient AEX context slots to hold another potential AEX after the re-entry. Without the possibility of a re-entry, in-enclave exception handling is effectively disabled. Making an enclave thread execution fully synchronous this way simplifies the reasoning about re-entrancy. However, this design choice limits the functionality of the enclave software. For example, the try-catch exception handling primitive widely used in modern high-level programming languages cannot leverage hardware exception support inside an enclave, making them inefficient and cumbersome to enable. It hinders the implementation of signal handling mechanisms commonly provided by modern OSes such as Linux, which are important to the functioning of user applications. Such limitations degrade the compatibility of Alibaba Inclavare and Fortanix Rust EDP with traditional programming models.

### 9.3 Re-entrant Exception Handling

An SGX runtime software may attempt to provide atomic primitives for re-entrant exception handling. One example is Graphene-SGX.

```
1  movq SGX_GPR_RIP(%rbx), %rax
2  leaq .Locall_about_to_eexit_begin(%rip), %r11
3  cmpq %r11, %rax
4  jb .Lhandle_interrupted_ocall_case_c
5  leaq .Locall_about_to_eexit_end(%rip), %r11
6  cmpq %r11, %rax
7  jae .Lhandle_interrupted_ocall_case_c
8
9  // ...
10
11 .Lhandle_interrupted_ocall_case_c:
12  movq %rdi, SGX_GPR_RSI(%rbx) # external event for .Lreturn_from_ocall
13  leaq .Lreturn_from_ocall_after_stack_restore(%rip), %rax
14  movq %rax, SGX_GPR_RIP(%rbx)
15  movq %rsi, SGX_GPR_RSP(%rbx)
16  movq $0, %gs:SGX_PRE_OCALL_STACK
17  andq $(~(RFLAGS_DF | RFLAGS_AC)), SGX_GPR_RFLAGS(%rbx)
18  jmp .Leexit_exception
```

**Listing 4: Emulation of part of the sanitization logic at enclave entry in the exception handler of Graphene-SGX.**

**Graphene-SGX.** It uses the same stack from the interrupted thread for exception handling. However, in our investigation, we find that Graphene-SGX does not blindly load the stack pointer from the SSA region. Instead, it examines the location of the AEX (the rip register value inside the SSA region), and handles it differently in different cases. For example, when Graphene-SGX finds that the AEX occurred within the sanitization logic at the enclave entry, it will emulate the unfinished sanitization logic. Instead of operating on real registers as in normal execution, it operates on the register values stored in the SSA region (see Listing 4). This separates the execution of the sanitization logic and the exception handler. Thus, when the enclave starts the post-sanitization processing of the AEX, the stack has already been correctly set up and is no longer controlled by the untrusted software.

```
1  leaq .Ltmp_rip_saved0(%rip), %rax
2  cmpq %rax, SGX_GPR_RIP(%rbx)
3  je .Lemulate_tmp_rip_saved0
4
5  leaq .Ltmp_rip_saved1(%rip), %rax
6  cmpq %rax, SGX_GPR_RIP(%rbx)
7  je .Lemulate_tmp_rip_saved1
8
9  leaq .Ltmp_rip_saved2(%rip), %rax
10 cmpq %rax, SGX_GPR_RIP(%rbx)
11 je .Lemulate_tmp_rip_saved2
12
13 jmp .Lemulate_tmp_rip_end
14
15 .Lemulate_tmp_rip_saved0:
16  # emulate movq SGX_CPU_CONTEXT_R15 - SGX_CPU_CONTEXT_RIP(%rsp), %r15
17  movq SGX_GPR_RSP(%rbx), %rax
18  movq SGX_CPU_CONTEXT_R15 - SGX_CPU_CONTEXT_RIP(%rax), %rax
19  movq %rax, SGX_GPR_R15(%rbx)
20 .Lemulate_tmp_rip_saved1:
21  # emulate movq SGX_CPU_CONTEXT_RSP - SGX_CPU_CONTEXT_RIP(%rsp), %rsp
22  movq SGX_GPR_RSP(%rbx), %rax
23  movq SGX_CPU_CONTEXT_RSP - SGX_CPU_CONTEXT_RIP(%rax), %rax
24  movq %rax, SGX_GPR_RSP(%rbx)
25 .Lemulate_tmp_rip_saved2:
26  # emulate jmp *%gs:SGX_TMP_RIP
27  movq %gs:SGX_TMP_RIP, %rax
28  movq %rax, SGX_GPR_RIP(%rbx)
29 .Lemulate_tmp_rip_end:
30  movq SGX_GPR_RSP(%rbx), %rsi
31  // ...
```

**Listing 5: Emulation of part of the enclave context restoration code in the exception handler of Graphene-SGX.**

```
1 .Ltmp_rip_saved0:
2  movq SGX_CPU_CONTEXT_R15 - SGX_CPU_CONTEXT_RIP(%rsp), %r15
3 .Ltmp_rip_saved1:
4  movq SGX_CPU_CONTEXT_RSP - SGX_CPU_CONTEXT_RIP(%rsp), %rsp
5 .Ltmp_rip_saved2:
6  jmp *%gs:SGX_TMP_RIP
```

**Listing 6: Part of the enclave context restoration code. Graphene-SGX emulates its execution instruction by instruction in the exception handler (see Listing 5).**

Besides emulating the sanitization logic and register setup at the enclave entry point, Graphene-SGX emulates the execution of the interrupted thread whenever the AEX occurs in other critical regions where the enclave state is unsafe for exception handling. Listings 5 and 6 show more examples. After the emulation, the AEX appears to have occurred outside the critical regions. By doing this, Graphene-SGX effectively makes the enclave exception handler safely re-entrant.

This design is significantly more nuanced and complex. We note that it has been the result of years of patching and revising. The signal and exception handling design for Graphene-SGX has undergone several iterations in the past three years [20, 24, 25].

**Alternative Implementation Strategies.** The design adopted by Graphene-SGX is not the only possible implementation strategy and can be generalized by addressing two key questions.

*Q1.* How can the enclave identify whether an exception occurred inside a critical section?

*Q2.* What should the enclave do when the untrusted OS attempts to deliver an exception during a critical section to the enclave?

**Tracking Critical Sections.** The enclave can track its own critical sections explicitly. Specifically, the enclave runtime software can maintain the location information (e.g., code address ranges) about all its critical sections. When required, it can check if the code address where the exception occurred falls within the range of known critical sections. This is the option chosen by Graphene-SGX. Alternatively, the enclave software can maintain a per-thread 1-bit flag in its *private* memory. It sets the flag whenever the enclave is about to enter a critical section, and clears it immediately after exiting a critical section. To check if an exception happened in a critical section, the enclave exception handler can check this flag.

**Handling Exceptions in Critical Sections.** If the enclave is interrupted midway in a critical section, one approach is to emulate the rest of the critical section. More precisely, the enclave exception handler can identify the point of interruption, look up the critical section, and emulate the remaining part of the interrupted critical section. After that, the handler can perform the real (non-emulated) exception handling. With this mechanism, the enclave gets an illusion that the exception occurred immediately after the critical section. This design works only in cases where the enclave runtime has sufficient information about the critical sections to emulate it. This is the option chosen by Graphene-SGX.

A second way is to postpone the exception handling. Instead of immediately invoking the exception handler, the enclave runtime software can choose to execute the current critical section. Once the section ends, the runtime executes the handler. This mechanism requires the enclave runtime to maintain the received exceptions (e.g., via setting a per-thread pending flag) and add logic at the end of each critical section to handle pending exceptions.

Finally, a straightforward way is to ignore any exceptions that arrive when a critical section is being executed. However, it is important to ensure functional correctness when the OS is cooperative. For instance, an exception should not be lost when a cooperative OS delivers it while the enclave is in a critical section, unless the enclave exposes sufficient information (e.g., by setting OS-visible critical section flags) to allow the OS to avoid delivering exceptions during critical sections.

**Caveats.** Although the above design options are conceptually simple, there are several implementation details that need careful attention. The first issue arises when an enclave has to maintain a data structure to track pending exceptions (e.g., a bitmap that records the postponed exception types). After the critical section, the enclave needs to read such data structures to process the postponed exceptions. Should the code that does this be included as part of the same critical section? If it is part of the critical section, the data structure operations must be made re-entrant. This is because, in a critical section, a delivered exception will trigger write operations to the data structures. If it is outside the critical section, one must either ensure the same re-entrancy property or ensure that no exception handling code contains critical sections, and hence may involve write operations to the data structures.

The second issue concerns the use of a critical flag. One must ensure that the flag covers all locations where exceptions should not be handled. One location particularly prone to negligence is immediately after an enclave entry (EENTER). As demonstrated by SmashEx, exceptions immediately after the enclave entry, when untrusted OS still controls the register values, cannot be handled directly. If the enclave relies on an instruction to set the flag after the enclave entry, this leave a window of time between enclave entry and when the flag is set. To avoid this problem, an enclave must ensure that the flag is set before an EEXIT.

A third issue stems from the requirement of exception-free handler implementations. Although one can carefully implement handler logic to be free of certain programming-oriented exceptions (e.g., divide by zero), OS-induced page faults are difficult to avoid. For instance, if an enclave uses custom page fault handlers on SGX2, delivery of page faults to the enclave cannot be delayed or ignored, especially for faults on pages accessed within a critical section of the exception handler itself.

In summary, while there are software-based strategies for achieving re-entrant exception handling, they introduce considerable complexity to ensure desired functionality and security.

### 9.4 Impact of Other Memory Defenses

A second line of defenses aim at thwarting the code-reuse attack steps (Steps 5 and 6) of SmashEx.

**Bypassing ASLR.** The SmashEx attack requires the attacker to know the exact address of the anchor. Since the OS allocates the virtual memory range for the enclave and sets the page permissions, the attacker knows that the enclave stack will be within a certain range. However, the enclave may randomize the base address of its stack (e.g., Asylo [31]) to prevent the attacker from predicting the anchor location accurately. The attacker can adapt SmashEx in the following ways to overcome this hurdle.

The first strategy follows the observation that Steps 1–4 are to overwrite more than one memory location. For example, in our Google Asylo [31] PoC exploit (elided here), we can overwrite 152 bytes, out of which 64 bytes are freely controllable by the attacker. The attacker can therefore set all those locations to the desired value for the anchor in the hope of hitting the actual anchor. Given that Google Asylo initializes the stack base address by advancing the stack by a random amount between 1 and 2048, this strategy has an attack success rate of 3.125% per trial.

One of the issues with using ASLR defenses in our context is that a failed trial results in an invalid memory access which in turn creates another exception. This gives the attacker additional opportunities to reenter the enclave. To concretely illustrate the issue, we implemented a multi-round proof-of-concept attack variant of SmashEx specialized for Google Asylo. A multi-round attack trial has multiple rounds, where each round executes Steps 1–4 of the SmashEx procedure to corrupt one location. Note that the attacker-corrupted location is used in Step 5. If we mispredict the anchor address as the corruption value, the enclave will potentially crash in the subsequent steps. What remains, therefore, is to keep iterating Steps 1–4 while making sure that the enclave does not progress to Step 5, i.e., to resume the return from ocall. To achieve this, the attacker uses an invalid `ecall` number[6] to enter the enclave in Step 1, instead of the one that requests a return from an `ocall`. When the enclave resumes execution after AEX at the entry point, it checks the `ecall` command. Since the `ecall` command is invalid, the enclave forces an EEXIT. As a result, Steps 5–6 will not take place and instead Steps 1–4 repeat. The attacker can then keep repeating this procedure of using bad `ecall` numbers and corrupt one new location each time reliably. Finally, after controlling sufficient locations, the attacker performs an EENTER with the correct `ecall` number. In this last iteration, the enclave performs Steps 5–6 and uses a bad stack value, thus corrupting the anchor reliably even in the presence of ASLR.

**Bypassing Stack Canary.** The stack canary is a widely-deployed defense against buffer overflow exploits [39]. The attacker has to corrupt unintended stack locations as a side effect of the memory corruption. While SmashEx does not involve a buffer overflow, it does corrupt unintended locations beyond the anchor itself. Therefore, it is conceptually possible to mitigate SmashEx with stack canaries. However, the stack canary supported in common modern C compilers (e.g., with `-fstack-protector-all` in GCC) does not help protect against SmashEx. Unlike the return address of a function call, the stack canary automatically generated by the compiler does not protect the saved `ocall` context, and hence the anchor, that SmashEx aims to control. Moreover, even if all code pointers on the stack have been carefully protected by stack canaries, the attacker can adapt SmashEx to launch a data-oriented attack [43] without controlling code pointers. In addition, due to a lack of checks on the stack pointer, on some SGX runtimes (e.g., Open Enclave) SmashEx has the option to control non-stack locations, including where the secret stack canary value is stored. This makes existing stack canary defenses ineffective against SmashEx.

## 9.5 Better Hardware Support for Atomicity

While it is possible to implement the enclave software in a safely re-entrant way, doing this entails a fairly complicated design which is difficult to reason about. This motivates us to propose strategies of enabling atomicity support in hardware, which SGX currently lacks. We start by examining the atomicity support available to the OS and traditional user applications.

**Atomicity in the OS.** Since the OS can configure hardware interrupt and exception sources (e.g., interrupt controllers), it can simply disable interrupts and exceptions whenever it desires atomicity. For SGX enclave software, however, an untrusted party (i.e., the OS) can trigger an exception at any time. The enclave has no way of controlling or predicting when it will be interrupted.

**Atomicity in Traditional User Applications.** Traditional user applications rely on the OS to control when they can be interrupted and re-entered in the midst of their execution (e.g., for signal handling). For example, POSIX-compliant OSes define the set of scenarios where they can deliver signals to a user process [22]. User processes can use the `sigprocmask` system call to dynamically enable or disable the delivery of a certain signal during runtime. For SGX enclaves, the OS or the host process is still in charge of invoking in-enclave exception handlers, but it is not trusted and should not be relied on to decide when to perform a re-entry.

**Enabling Atomicity Primitives in Hardware.** We point out that in both the above cases, the atomicity guarantee is provided by a different but trusted entity (the hardware or the OS) through disabling either interrupts or re-entry upon an interrupt. For an SGX enclave, only the hardware can be such a trusted entity. We discuss the potential changes in the SGX hardware abstraction to enable atomicity primitive for enclaves. Following the inspiration from the atomicity primitives available to the OS and traditional user applications, we discuss two directions: disabling exceptions and disabling re-entry.

*Direction 1: Temporarily Disabling Interrupts and Exceptions.* Intel SGX can be adapted to allow enclaves to dynamically enable or disable interrupts and exceptions from hardware, similarly to the primitives OSes use to achieve atomicity. The SGX hardware can protect the enclave execution from being interrupted at the request of the enclave. A naïve design that allows the enclave to use this primitive without restrictions will enable an enclave to fully occupy a hardware thread for an arbitrarily long period, thus launching denial-of-service (DoS) attacks against the OS. To avoid DoS attacks, the SGX hardware can let the OS decide whether to accept or deny such requests from enclaves. The hardware relays the decision of the OS to the enclave, who in turn can make an informed decision about executing critical code. For example, the OS may base such decisions on a pre-exchanged quota for interrupt disabling: it may permit the enclave to run with interrupts disabled for (say) 100 cycles in every 10K cycles executed in the enclave. In such a case, corresponding support for counting and limiting the enclave execution cycles will need to be available inside the hardware. Such an enclave-OS contract facilitated and enforced by the hardware, if designed carefully, can guarantee atomicity while preventing DoS.

*Direction 2: Temporarily Disabling Enclave Re-entry.* Instead of blocking interrupts or exceptions, another option is to allow the enclave to disable enclave re-entry during runtime. In this design, the enclave can still be interrupted when it has disabled enclave re-entry, leaving untrusted software the chance to perform exception handling and manage resources accordingly. However, the SGX hardware only allows it to resume the enclave execution via ERESUME, but disallows re-entry into the enclave via EENTER. Immediately after enclave entry, since the enclave software needs to perform crucial operations to complete a context switch into the enclave, the enclave hardware should preferably automatically mask enclave re-entry by default to ensure its atomicity.

Both directions pose the risk of opening a new side channel. The attacker may learn whether an enclave is inside a critical code region simply by attempting to deliver an exception into it. Nevertheless, we believe addressing atomicity on the OS-enclave interface through a carefully designed hardware abstraction is promising future work. We hope these directions offer a starting point.

## 10 RELATED WORK

The SmashEx attack is targeted at Intel SGX enclaves. It stems from unsafe re-entrancy at the OS-enclave interface. We have demonstrated that, if exploited, it can lead to code-reuse attacks. In this section we examine prominent work on securing host-enclave interfaces on Intel SGX, code-reuse attacks targeting SGX enclaves, and re-entrancy vulnerabilities in non-enclave settings.

---
[6]The `ecall` number is an integer that the untrusted software passes to the enclave upon an `ecall` to indicate which enclave function to execute.

**Security of Synchronous SGX Interfaces.** Since the introduction of Intel SGX, there has been abundant work on the security of the synchronous interfaces between untrusted software and SGX enclaves. Previous work has discovered that an attacker can compromise the confidentiality and integrity of an enclave by providing malicious system call return values, referred to as Iago attacks [36]. Eliminating such threats requires enclave software to carefully scrutinize system call return values passed into an enclave [30, 57, 60], with the aid of formal verification [58] or software testing techniques [40]. In addition, enclave runtimes may forget to clean certain registers after a context switch into an enclave, thus opening up the enclave to attacks [29, 61]. The synchronous interface has been a subject of comprehensive survey and categorization of attacks [44]. Unlike these lines of work, our paper examines the security of asynchronous OS-enclave interfaces.

**Security of Asynchronous SGX Interfaces.** Existing attacks have shown that timer interrupts or page faults can be leveraged to leak enclave secrets through side channels [34, 62, 65, 66]. Defending against such side-channel attacks is non-trivial [37, 55, 56]. Previous work has examined the attack avenue of enclave thread scheduling. In the SGX threat model, the attacker can control the scheduling and influence the enclave logic. Such manipulations can compromise enclave confidentiality and integrity if the enclave logic is influenced by scheduling. For example, the attacker can affect the enclave behavior by exploiting existing synchronization bugs [2] or breaking assumptions made by the enclave application regarding the thread scheduling algorithm [63]. However, the security implications of the asynchronous interfaces of SGX enclaves have not been comprehensively studied. Specifically, to our knowledge, our work is the first to study the security implications at the OS-enclave interface for asynchronous exceptions on SGX.

**Code-reuse Attacks on SGX.** The prevalence of code-reuse attacks in non-enclave applications is well studied [59]. Although enclaves reduce the size of the trusted computing base, they are susceptible to corruption if the enclave code has unsafe memory usage. Thus, enclaves are not immune to code-reuse attacks [38]. Dark-ROP [45] demonstrates a ROP attack even when the enclave binary is end-to-end encrypted [32, 53] such that the attacker cannot inspect it. They assume a fixed enclave address space layout, which allows the attacker to probe the locations of useful gadgets through trials-and-errors. This assumption is justified by the difficulty in applying defense techniques such as ASLR to SGX enclaves due to the constraints imposed by the Intel SGX design. SGX-Shield [54] proposes a strategy to enable ASLR in SGX enclaves and prevent code-reuse attacks. However, as shown in the subsequent work, SGX-Shield does not randomize the code inside the trusted runtime of the enclave. This allows the attacker to exploit memory-unsafe enclave code and launch powerful ROP attacks [33]. SmashEx demonstrates a code-reuse attack on enclaves. However, unlike the existing work, we do not assume a pre-existing memory vulnerability in the enclave software.

**Re-entrancy Vulnerabilities & Defenses.** Traditional asynchronous interfaces, such as the signal handler, are prone to re-entrancy challenges [67]. Such vulnerabilities are common in several other systems [5–7, 10, 11, 42, 46]. SmashEx is the first attack that exploits re-entrancy vulnerabilities in the context of Intel SGX. Preventing re-entrancy bugs in general involves introducing a notion of atomicity. For instance, when the code is operating in a critical section, the user application can request the OS to mask certain signals (i.e., to pause their delivery) [27]. Our work makes the first attempt to compare and contrast exception handling in Intel SGX versus traditional systems. Our findings highlight the need for better hardware abstractions to enable safely re-entrant enclave code.

SmashEx brings attention to a new avenue of powerful attacks on Intel SGX. It can serve as a motivation to further scrutinize and fortify the enclave asynchronous interface.

## 11 RESPONSIBLE DISCLOSURE

We informed the affected parties—Intel for Intel SGX SDK and Microsoft for Open Enclave SDK—on 3 May 2021. Intel and Microsoft acknowledged the attack and assigned CVE-2021-0186 and CVE-2021-33767 respectively [8, 9]. After due process, Intel and Microsoft released patches for SmashEx on 13 July 2021. In addition, they released public advisories on 13 July 2021 [21] and 12 October 2021 [19]. Further, we have assisted Intel and Microsoft to coordinate responsible disclosures to the affected runtimes listed in Table 1, where it was requested.

## 12 CONCLUSION

Asynchronous exception handling is a commodity functionality for real-world applications today, which are increasingly utilizing enclaves. In this work, we show the importance of providing atomicity guarantees at the OS-enclave interface for such exceptions. We have introduced a new attack called SmashEx in this work, which exploits the inherent re-entrancy interface required in exception handling on SGX. Our exploits demonstrate the issue concretely on popular SGX runtime frameworks. We hope our work initiates careful consideration for asynchronous exception handling in existing SGX frameworks as well as in future enclave designs.

## AVAILABILITY

We maintain further information regarding SmashEx, including how to acquire the proof-of-concept exploits for educational purposes, at https://jasonyu1996.github.io/SmashEx/.

## ACKNOWLEDGMENTS

We thank the anonymous CCS reviewers for their valuable suggestions. We also thank Ivan Puddu for help with SGX-Step. This work was supported by Crystal Center at National University of Singapore. Zhiping Cai's work was funded by National Natural Science Foundation of China (62072465). Any opinions, findings, and conclusions or recommendations expressed in this material are those of the authors only.